# Lasing oscillation in a three-dimensional photonic crystal nanocavity with a complete bandgap

Aniwat Tandaechanurat*, Satomi Ishida, Denis Guimard, Masahiro Nomura, Satoshi Iwamoto and Yasuhiko Arakawa*

*Institute for Nano Quantum Information Electronics, The University of Tokyo, 4-6-1 Komaba, Meguro, Tokyo 153-8505, Japan*

*e-mail: aniwat@iis.u-tokyo.ac.jp; arakawa@iis.u-tokyo.ac.jp

**Photonic crystals[1,2] have been extensively used to control and manipulate photons in engineered electromagnetic environments provided by photonic bandgap (PBG) effects, which are a key tool for realizing future optoelectronic devices, such as highly efficient lasers. So far, photonic-crystal-cavity-based lasers have exclusively been demonstrated in two-dimensional (2D) photonic crystal geometries[3-6]. However, the full confinement of photons and control of their interaction with materials can only be achieved with the use of 3D photonic crystals possessing complete PBGs[7-16]. Here, we demonstrate, for the first time, the realization of lasing oscillation in a 3D photonic crystal nanocavity. The laser is realized by coupling a cavity mode, which is localized in a complete PBG and exhibits the highest quality factor ($Q$ factor) of ~38,500, with semiconductor quantum dots (QDs). We show a systematic change in the laser characteristics, including the threshold and the spontaneous emission coupling factor ($\beta$) by controlling the crystal size, which consequently changes the strength of photon confinement in the third dimension. Our achievement opens a new and practical pathway for exploring the physics of light-matter interactions in a nanocavity-single-QD coupling system, where both photons and electrons are three-dimensionally confined, as well as for realizing 3D integrated photonic circuits.**

3D photonic crystals provide the capability to inhibit light propagation in any direction through their complete PBGs. By introducing an artificial defect into a perfect crystal, 3D photonic crystal nanocavity modes with high $Q$ factors and small mode volumes of the order of a cubic wavelength can be localized in a complete PBG, which is crucial for realizing ultimate control of light-matter interaction. For example, lasers constructed using 3D photonic crystal cavities offer many advantages over those using 2D structures because of the additional dimensionality of the photon confinement. In principle, 3D photonic crystal cavity lasers are expected to consume considerably less power to be switched on than that consumed in 2D systems because of a stronger suppression of undesired spontaneous emission[1]. In addition, 3D photonic crystals also offer the possibility to directly funnel the light output from the cavity[17] and subsequently guide it through optical circuit paths to other optical devices on the same chip with low loss[8,13]. Further, 3D photonic crystals can be made to couple with material with the gain polarized in any direction because of the omnidirectional property of the complete PBG[1,18]. Despite these striking advantages, the realization of 3D photonic crystal lasers has been hindered by difficulties in the fabrication of intricate 3D structures that possess a complete PBG and contain a high-$Q$ cavity incorporated with an efficient light-emitting material. Lasing in 3D photonic crystals reported to date has usually been achieved by using a bandedge mode[19-21] or an unintentionally created defect mode[22] associated with a pseudogap, but not the complete PBG. Therefore, it would be difficult to use such structures to gain the above-mentioned advantages because they can manipulate photons only in a specific direction. Although several efforts have been devoted to developing the quality of active cavities embedded in complete-PBG crystals[14-16], the lasing oscillation has yet to be realized owing to a relatively low $Q$ factor and/or insufficient material gain. Here, we report the fabrication of a 3D photonic crystal nanocavity with a high-$Q$ cavity mode, which is localized in a complete PBG and exhibits a record $Q$ factor of 38,500, by using micromanipulation techniques[15,16]. Coupling this high-$Q$ mode with high-quality indium arsenide (InAs) QDs[23], we demonstrate the first 3D photonic crystal nanocavity laser under a pulsed operation at a low temperature.

A schematic representation of the fabricated 3D photonic crystal, a so-called woodpile structure, is shown in Fig. 1a. The structure is a stack of gallium arsenide (GaAs)-based 2D thin layers, each containing a line-and-space pattern with eleven in-plane ($x$-$y$ plane) rods, fabricated using the micromanipulation techniques. An active layer embedding a point-defect structure with the dimension of 1.15 μm × 1.15 μm and three-layer stacked InAs QD layers (see Fig. 1b) with a dot density of ~4 × $10^{10}$ cm$^{-2}$ per layer as a cavity and a light emitter, respectively, was sandwiched between the upper and the lower layers. The QD ground state emission peak was at 1.26 μm at 7 K. The in-plane periodicity, thickness, and width of rods of each layer, including the active layer, were set to 500 nm, 150 nm, and 130 nm, respectively. The number of lower layers of the structure was set to be twelve, whereas that of the upper layers was increased from an initial value of two to twelve. Scanning electron microscope (SEM) images of the fabricated structure are shown in Fig. 1c. Rectangular posts seen in the image help to guide the layers to a designated position, allowing a high-precision assembly with a stacking error of less than 50 nm. The fabrication time required for assembling one layer was less than 10 min.

As further upper layers were stacked onto the structure, photoluminescence (PL) measurements were performed at 7 K. We initially investigated the effect of the number of upper layers on the $Q$ factor of a high-$Q$ cavity mode. Figure 2a shows the PL spectrum for the structure with six upper layers, exhibiting a number of cavity modes in a complete PBG between 1085 and 1335 nm, predicted by calculations using a plane-wave expansion (PWE) method. The $Q$ factor of the mode with the most pronounced intensity at approximately 1200 nm, very close to the centre wavelength of the PBG where the localization of photons was the strongest[16], is plotted as a function of the number of upper layers in Fig. 2b. We estimated the value of the measured $Q$ from the Lorentzian fit to the experimental data measured at the transparency pump power. The $Q$ factor grew exponentially with the number of upper layers as a result of the strengthening of the PBG effect[24] and reached the value of 38,500 (Fig. 2c), which is the highest value reported so far for 3D photonic crystal nanocavities, when the number of upper layers was twelve. These results agreed well with the theoretical calculations. In addition to the achievement of the highest $Q$ factor, these results also present experimental evidence that confirms the direct control of the photon confinement in the third dimension. The tendency of $Q$ started to saturate when the number of upper layers was more than eight, where it approached the limit value mainly determined by the small size of the photonic crystal in the in-plane direction. Thus, increasing the number of rods in the in-plane direction could further improve the *Q* factor.

A pulsed operation of 3D photonic crystal lasers was achieved when the number of upper layers was six or more. The structure was pumped by 8-ns pulses of 905-nm wavelength from a semiconductor laser with a repetition rate at 25 kHz. Figure 3 shows the output power of the lasing mode as a function of the pump power (*L-L* plot) for the laser structure with different numbers of upper layers. The lasing spectrum for each structure at averaged pump power of 30 μW is shown in the inset. The lasing thresholds were obtained by extrapolating back the plots from above thresholds to zero output power. The averaged lasing threshold was reduced from 2.4 μW to 2 and 1.3 μW when the number of upper layers was increased from six to eight and twelve, respectively, which could be attributed to the improvement of *Q*. We simulated the values of the threshold power as a function of *Q* factor using coupled rate equations for the carrier density and the photon density[25]. The general tendency of the simulated threshold power was in good agreement with that of the experimental values as shown in Fig. 4a. In addition, we also observed a reduction in the laser slope efficiency, in which light output radiated out from the cavity into the top side of the structure was more strongly suppressed as more upper layers were added, consistent with the extraction efficiency calculated by comparing the cavity loss rate into the top side with the total loss rate defined as the sum of the cavity loss rate into all directions and the scattering loss rate due to fabrication imperfections. We then evaluated the value of the spontaneous emission coupling factor *β*, which is a measure of the coupling efficiency of the spontaneous emission to the lasing mode. Because of the existence of the complete PBG in the present laser, where the lasing mode was strongly confined and well isolated from other leaky optical modes, a high value of *β* was expected. Fitting the experimental *L-L* plots with steady-state solutions of the rate equations on double logarithmic scales as shown in Fig. 4b, we determined $\beta$ of 0.54, 0.67, and 0.92 for the laser structure with six, eight, and twelve upper layers, respectively. In particular, for the structure with twelve upper layers the value of $\beta$ was very close to the theoretical limit of unity[26], attributing to the very high $Q$ factor of the lasing mode localized in the complete PBG.

We have demonstrated lasing oscillation in a 3D photonic crystal nanocavity by coupling a high-$Q$ cavity mode with QDs. Localizing the lasing mode in a complete PBG allows us to strongly confine light in the cavity and to effectively restrict undesired spontaneous emission; these were evident from the high cavity $Q$ and the high $\beta$ factor, respectively. We believe that our demonstration of lasers in 3D photonic crystals should advance the development of practical 3D integrated photonic circuits significantly as now the light source has already been provided. Furthermore, introducing a single QD into the present high-$Q$ 3D photonic crystal nanocavity would establish an ideal solid-state system for the study of light-matter interaction between three-dimensionally confined photons and electrons enclosed in a completely engineered electromagnetic environment. This may lead us to the discovery of new physics in the field of cavity quantum electrodynamics.

**Figure 1** (a) Schematic illustration of a fabricated 3D photonic crystal. A portion of the upper layers is removed to show the cross section of the stacked structure and to reveal the cavity structure. (b) Illustration of a cross section of an active layer showing three-layer stacked QDs (upper). The lower figure is a $1 \times 1$ $\mu m^2$ planar atomic force microscope image of InAs QDs on a GaAs matrix. (c) SEM images of the fabricated 25-layer woodpile structure shown in bird's eye (left) and top (right) views.

**Figure 2** (a) PL spectrum for the structure with six upper layers pumped with a continuous-wave titanium-sapphire laser. The complete PBG is shown as an unshaded area. A dotted line represents the centre wavelength of the PBG. (b) Dependence of the measured $Q$ factor of the mode at 1200 nm on the number of upper layers. A line is the exponential fit. The insets show SEM images of the fabricated structures with the corresponding number of upper layers. (c) High-resolution PL spectrum fitted with a Lorentzian function for the structure with twelve upper layers, exhibiting the highest $Q$ factor of approximately 38,500.

**Figure 3** *L-L* plots for the laser structure with different numbers of upper layers. Lines are the linear fits for the experimental plots above the thresholds. Inset, lasing spectrum for each structure at an averaged pump power of 30 $\mu$W.

**Figure 4** (a) Threshold power (circles) and laser slope efficiency (squares) as a function of the $Q$ factor. The $Q$ factors are 7,400, 18,300, and 38,500 for the structure with six, eight, and twelve upper layers, respectively (see also Fig. 2b). Also plotted are the simulated threshold power (solid curve) calculated using rate equations and the light extraction efficiency (dashed curve) as a function of the $Q$ factor. (b) Experimental *L-L* plots (symbols) and fitting curves obtained by the rate equation analysis with best-fitted values of *β*. The plot and curve for the structure with six upper layers have been shifted upwards to improve the clarity of the graph.

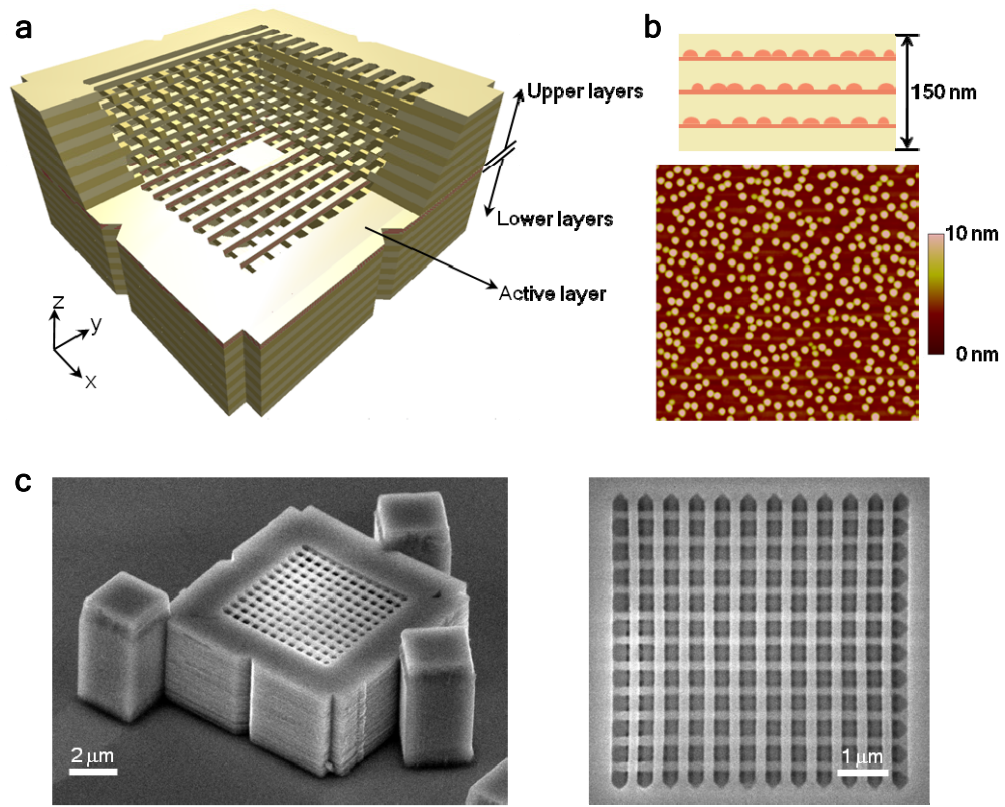

**a**

Upper layers

Lower layers

Active layer

z
y
x

**b**

150 nm

10 nm

0 nm

**c**

2 µm

1 µm

**Fig. 1**

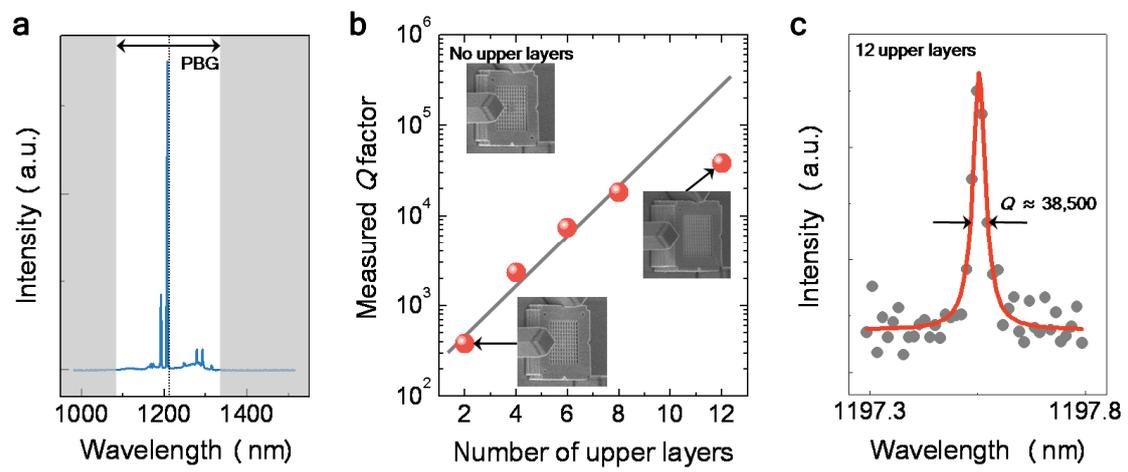

**a**

Intensity（a.u.）

PBG

1000　1200　1400
Wavelength（nm）

**b**

No upper layers

Measured $Q$ factor

$10^6$
$10^5$
$10^4$
$10^3$
$10^2$

2　4　6　8　10　12
Number of upper layers

**c**

12 upper layers

Intensity（a.u.）

$Q \approx 38,500$

1197.3　1197.8
Wavelength（nm）

**Fig. 2**

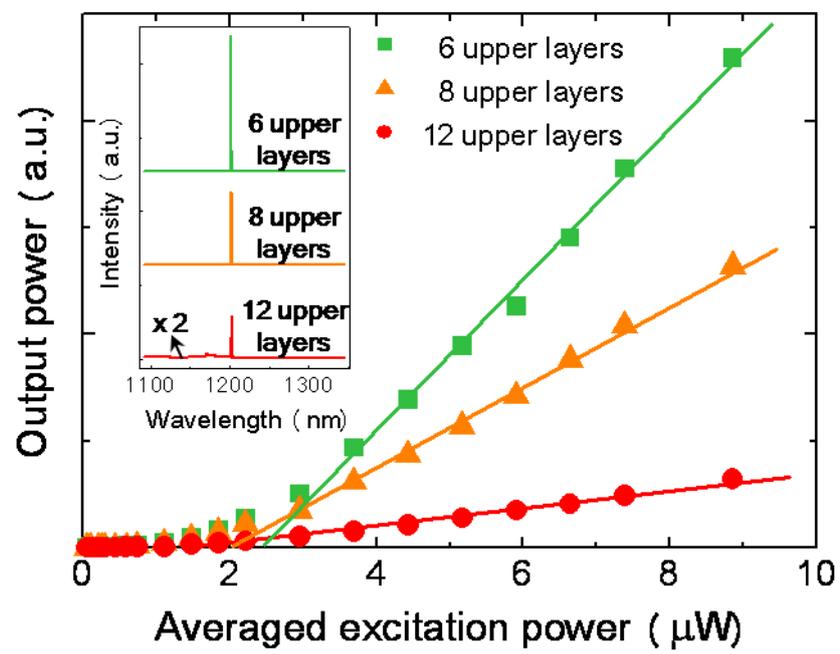

**Fig. 3**

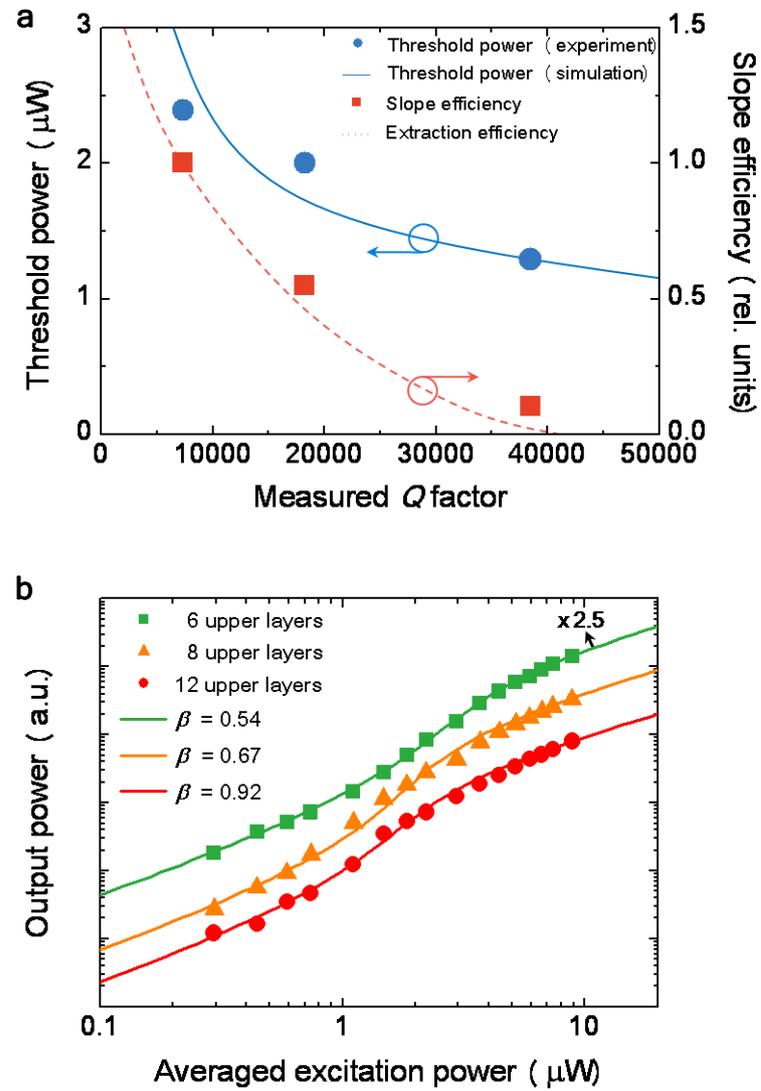

**Fig. 4**